\documentclass[twocolumn,trackchanges,twocolappendix]{aastex702}

\begin{document}

\title{The Common Envelope Evolution Outcome. III. the Improvement of Stellar Binding Energy with the Envelope Residual}

\author[orcid=0009-0001-3638-3133]{Lifu Zhang}
\affiliation{Yunnan Observatories, Chinese Academy of Sciences, Kunming 650216, People's Republic of China}
\affiliation{International Centre of Supernovae (ICESUN), Yunnan Key Laboratory of Supernova Research, Kunming 650216, P.R. China}
\affiliation{University of Chinese Academy of Sciences, Beijing 100049, People's Republic of China}
\email{zhanglifu@ynao.ac.cn}

\author[orcid=0000-0002-6398-0195]{Hongwei Ge}
\affiliation{Yunnan Observatories, Chinese Academy of Sciences, Kunming 650216, People's Republic of China}
\affiliation{International Centre of Supernovae (ICESUN), Yunnan Key Laboratory of Supernova Research, Kunming 650216, P.R. China}
\affiliation{University of Chinese Academy of Sciences, Beijing 100049, People's Republic of China}
\email[show]{gehw@ynao.ac.cn}

\author[orcid=0009-0006-1130-9777]{Dylan Hebrail}
\affiliation{Lund Observatory, Division of Astrophysics, Department of Physics, Lund University, Box 118, 22100 Lund, Sweden}
\email{dylan.hebrail@fysik.lu.se}

\author[orcid=0000-0001-9204-0779]{Ross Church}
\affiliation{Lund Observatory, Division of Astrophysics, Department of Physics, Lund University, Box 118, 22100 Lund, Sweden}
\email{ross.church@fysik.lu.se}

\author[orcid=0009-0000-6595-2537]{Mingkuan Yang}
\affiliation{Yunnan Observatories, Chinese Academy of Sciences, Kunming 650216, People's Republic of China}
\affiliation{Key Laboratory of Optical Astronomy, National Astronomical Observatories, Chinese Academy of Sciences, Beijing 100101, China}
\email{yangmk@bao.ac.cn}

\author[orcid=0000-0002-1421-4427]{Zhenwei Li}
\affiliation{Yunnan Observatories, Chinese Academy of Sciences, Kunming 650216, People's Republic of China}
\affiliation{International Centre of Supernovae (ICESUN), Yunnan Key Laboratory of Supernova Research, Kunming 650216, P.R. China}
\affiliation{University of Chinese Academy of Sciences, Beijing 100049, People's Republic of China}
\email{lizw@ynao.ac.cn}

\author[orcid=0009-0006-9211-2860]{Hailiang Chen}
\affiliation{Yunnan Observatories, Chinese Academy of Sciences, Kunming 650216, People's Republic of China}
\affiliation{International Centre of Supernovae (ICESUN), Yunnan Key Laboratory of Supernova Research, Kunming 650216, P.R. China}
\affiliation{University of Chinese Academy of Sciences, Beijing 100049, People's Republic of China}
\email{chenhl@ynao.ac.cn}

\author[orcid=0000-0003-4265-7783]{Dengkai Jiang}
\affiliation{Yunnan Observatories, Chinese Academy of Sciences, Kunming 650216, People's Republic of China}
\affiliation{International Centre of Supernovae (ICESUN), Yunnan Key Laboratory of Supernova Research, Kunming 650216, P.R. China}
\affiliation{University of Chinese Academy of Sciences, Beijing 100049, People's Republic of China}
\email{dengkai@ynao.ac.cn}

\author[orcid=0000-0001-5284-8001]{Xuefei Chen}
\affiliation{Yunnan Observatories, Chinese Academy of Sciences, Kunming 650216, People's Republic of China}
\affiliation{International Centre of Supernovae (ICESUN), Yunnan Key Laboratory of Supernova Research, Kunming 650216, P.R. China}
\affiliation{University of Chinese Academy of Sciences, Beijing 100049, People's Republic of China}
\email{cxf@ynao.ac.cn}

\author[orcid=0000-0001-9204-7778]{Zhanwen Han}
\affiliation{Yunnan Observatories, Chinese Academy of Sciences, Kunming 650216, People's Republic of China}
\affiliation{International Centre of Supernovae (ICESUN), Yunnan Key Laboratory of Supernova Research, Kunming 650216, P.R. China}
\affiliation{University of Chinese Academy of Sciences, Beijing 100049, People's Republic of China}
\email{zhanwenhan@ynao.ac.cn}


\begin{abstract}

Common-envelope evolution (CEE) is a key process in the evolution of close binary systems. Many important astrophysical objects and evolutionary stages are closely related to CEE, including white dwarf binaries, hot subdwarfs, and gravitational wave mergers.
In the standard energy formalism of CEE, the binding energy of the donor envelope plays a crucial role, as it directly affects the final orbital period after CEE and serves as a key physical parameter in binary population synthesis studies.
However, the currently adopted binding energy suffers from large uncertainties, mainly because the envelope binding energy of giant-branch stars varies strongly near the helium-core boundary. In addition, the expansion of the star during CEE can also affect the binding energy. 
To address these issues, we introduce an improved binding energy for the envelope mass residual. Based on adiabatic mass loss models, we recalculate the distribution of the CEE binding-energy parameter $\lambda$ for stars with different masses and at different evolutionary stages, and we analyse the effects of envelope mass residual and adiabatic expansion. Due to the envelope mass residual, the $\lambda$ of some donors can increase by one to two orders of magnitude at the late red giant branch and asymptotic giant branch stages. Furthermore, we provide interpolation grids and fitting formulae for these results, which can be readily applied to various binary population synthesis codes.

\end{abstract}



\section{Introduction}\label{sec-intro}

During binary evolution, a star may fill its Roche lobe \citep{1985ibs..book...39W,2023pbse.book.....T} and initiate Roche-lobe overflow (RLOF). If the mass transfer rate becomes sufficiently high that the transfer timescale is much shorter than the thermal timescale of the donor star \citep{1987ApJ...318..794H,1997A&A...327..620S}, the donor envelope expands adiabatically \citep{PaperI,PaperII,PaperIII} and eventually engulfs the binary orbit, leading to a common-envelope (CE) scenario \citep{1976IAUS...73...75P,1993PASP..105.1373I}. Common-envelope evolution (CEE) plays a critical role in the evolution of close binary systems \citep{2011ApJ...730...76I,2013A&ARv..21...59I}, particularly in the formation of short-period binaries. During this unstable mass-transfer phase, the two stellar cores spiral inward within the CE. In some cases, this spiral-in phase may eventually lead to a binary merger \citep{2006A&A...451..223T,2011A&A...528A.114T,2013Sci...339..433I}.

CEE influences the formation of many systems, including white dwarf binaries \citep{2026ApJ..1007...15S}, cataclysmic variables \citep{2015ApJ...809...80G,2023MNRAS.525.3597I}, hot subdwarf binaries \citep{2002MNRAS.336..449H,2009ARA&A..47..211H,2016PASP..128h2001H}, X-ray binaries \citep{2000ApJ...530L..93T,2012ApJ...756...85S,2025A&A...704A.219S,2023ApJ...945....7G}, and double compact objects \citep{2023hxga.book..129B}. Several key observational phenomena are also affected by CEE, such as type Ia/Ib supernovae \citep{2004MNRAS.350.1301H,2023RAA....23h2001L}, gravitational-wave sources \citep{2020ApJ...904...56G,2024ResPh..5907568L,2025A&A...695A.199K}, luminous red novae \citep{2011A&A...528A.114T}, and Thorne--\.Zytkow objects \citep{1975ApJ...199L..19T,2017ApJ...846..170T,2025A&A...694A..83N}. In recent years, a large number of post-common-envelope binaries (PCEBs) have been observed \citep{2010A&A...520A..86Z,2011A&A...536A..43N,2017MNRAS.470.1442C,2024MNRAS.52711719Y,2024PASP..136h4202Y}. However, the detailed physics of CEE remain poorly understood, and observational evidence for CE systems is still scarce. Three-dimensional hydrodynamical simulations have developed rapidly in recent years \citep{2016MNRAS.462..362I,2020A&A...644A..60S,2020A&A...642A..97K,2022MNRAS.512.5462L,2024ApJ...963L..35C}, but owing to computational efficiency and cost limitations, they still cannot cover the full range of CE systems.

Currently, the traditional CEE energy formalism \citep{1993PASP..105.1373I,2002MNRAS.336..449H} still plays an important role in binary evolution. A portion of the orbital energy is used to eject the CE. However, the energy sources driving CEE ejection remain uncertain. To unbind the envelope, the binding energy $E_\textrm{bind}$ of the donor's envelope is a key parameter \citep{1984nsf....8317916W,1990ApJ...358..189D,2022ApJ...933..137G,2024ApJ...961..202G}.

$E_\textrm{bind}$ is mainly determined by the gravitational potential energy and the internal energy. A few studies suggest that enthalpy may also contribute to CE ejection \citep{2013A&ARv..21...59I,2016RAA....16..126W}. Internal energy, which is the most uncertain component, includes thermal energy, radiation energy, ionization energy, and dissociation energy. Recent 1D and 3D CEE simulations suggest that hydrogen recombination energy may plays a significant role \citep{2020A&A...644A..60S,2022MNRAS.512.5462L,2024ApJ...963L..35C}. The contribution of other energy forms remains unclear, and most studies still treat internal energy as the primary component. Many previous works have provided $E_\textrm{bind}$ distributions \citep{2010ApJ...716..114X,2011ApJ...743...49L,2011MNRAS.411.2277D,2016RAA....16..126W}, which are widely used in binary population synthesis (BPS).

However, this energy prescription faces several unresolved issues. For some PCEBs, the observed systems require different ejection parameters to match the model predictions \citep{2026ApJ..1004...31L,2022MNRAS.513.3587Z,2023MNRAS.518.3966S,2024A&A...686A..61B,2024ApJ...961..202G}. Meanwhile, the final mass $M_\textrm{f}$ after CEE is a key assumption to determine $E_\textrm{bind}$. \cite{2011ApJ...730...76I} suggest using the boundary at the envelope's maximum compression layer. The most widely adopted assumption for $M_\textrm{f}$ is the edge of the helium core. From a fundamental physics perspective, the classical energy formalism has the following drawbacks:

\begin{itemize}
	\item [1)] Both observational and simulation studies indicate that a residual hydrogen envelope often remains. Examples include hot subdwarf \citep{2002MNRAS.336..449H} stars and low mass white dwarf binaries \citep{2018A&A...614A..49C,2018A&A...620A.196C}. Asteroseismic observations also support this conclusion \citep{2012A&A...539A..12F,2026arXiv260702720C}. Recent adiabatic simulations provide a self-consistent treatment of the residual envelope for specific binary systems \citep{2024ApJ...961..202G}. A thin envelope can significantly affect $E_\textrm{bind}$ at certain evolutionary stages, as we will show later.
	\item [2)] The classical $E_\textrm{bind}$ formula assumes that the post-CEE structure resembles the initial core profile of the donor. Thus, it is reasonable to use the initial donor structure to compute gravitational and internal energies. In reality, however, the boundary conditions of the star change during CEE mass loss. When the outer layers are removed, the interior expands. Given the short CEE timescale, this expansion is likely adiabatic \citep{2024ApJ...961..202G}. After CEE, the remaining star gradually returns to thermal equilibrium.
\end{itemize}

In this work, we address the two issues above. We adopt adiabatic mass loss models to investigate the effects of post-CE envelope residual and adiabatic expansion on $E_\textrm{bind}$. In Section\,\ref{sec:method}, we describe the adiabatic models and the improvement of detailed CEE energy formalism. Section\,\ref{sec:correction} presents the calculation results from initial structure before CEE and the adiabatically corrected $E_\textrm{bind}$. In Section\,\ref{sec:fit}, we offer the interpolation grids and fitting formulas based on previous calculations. Finally, in Section\,\ref{discuss}, we list the potential impacts of the results presented in this article.

\section{Models and methods} \label{sec:method}

In this section, we introduce the stellar models and calculation methods used in this work. In the first part, we briefly introduce the stellar evolution grid and the adiabatic mass loss models. In the second part, we provide a detailed description of CEE energy formalism. The expression of the improvement of envelope mass residual is also presented in the second subsection.

\subsection{Stellar Evolution Models}

The data are taken from a series of adiabatic mass loss models \citep{PaperI,PaperII,PaperIII}, which are constructed using the EZ version \citep{,2004PASP..116..699P} of the \textit{STARS} code \citep{1971MNRAS.151..351E,1972MNRAS.156..361E,1973MNRAS.163..279E}, a one-dimensional non-Lagrangian stellar evolution program. We adopt a Population I metallicity (solar metallicity) of $Z=0.02$. Other metallicities will be considered in future works. The overshooting parameter $\delta=0.12$ (\citealt{1997MNRAS.285..696S,1998MNRAS.298..525P}) and the mixing-length parameter $\alpha=2.0$ (\citealt{1998MNRAS.298..525P}) are both calibrated values. The stellar evolution grid covers masses from $0.1\,M_{\odot}$ to $100\,M_{\odot}$, and evolutionary stages from the main sequence (MS), through the Hertzsprung gap (HG) and the red giant branch (RGB), to the asymptotic giant branch (AGB). The grid points are uniformly spaced in $\log M$ and $\log R$ (base-10 logarithm, throughout this work). All models currently neglect stellar winds and rotation. Consequently, the stars have smaller cores but can reach larger radii, which helps to cover a wider parameter space in radius.

The adiabatic mass loss models are based on the grid described above. For each model, mass is removed from the envelope while the entropy profile remains fixed in the stellar mass coordinate. Adiabatic mass loss models can simulate the stellar structure in detail during rapid binary mass-transfer process, and it is currently of great improvement in studies of the stability of binary mass transfer. The binary mass-transfer criterion derived from adiabatic mass loss models is now widely applied to various stellar populations \citep{2019MNRAS.490.3740N,2023A&A...669A..82L,2024A&A...681A..31P,2026ApJ..1007...15S}. These models are also used to evaluate the CEE parameters of sdB binary systems \citep{2022ApJ...933..137G,2024ApJ...961..202G}.

The base of a giant branch is generally defined as the point where the convective-envelope mass accounts for one-third of the total stellar mass \citep{1998MNRAS.298..525P}. However, there exists a short transition phase near the base of the giant branch. These transition-phase models interfere with the trend of the binding energy along the giant branch.
For convenience, in this work we adopt a stricter definition for the base of the RGB (BRGB) and AGB (BAGB) stage, namely the points at which the convective-envelope mass reaches 50\% of the total envelope mass. The radius of BRGB stars approximately follows:
\begin{equation}\label{eq_BRGB}
\begin{aligned}
&\log \left ( \frac{R_\textrm{BRGB}}{R_\odot} \right )= \\
&-0.502\left [ M_\textrm{i} \right ] ^3+1.075\left [ M_\textrm{i} \right ] ^2+1.396\left [ M_\textrm{i} \right ] +0.364, \\
&\textrm{with } \left [ M_\textrm{i} \right ]=\log \left ( \frac{M_\textrm{i}}{M_\odot} \right ),  0.3<M_\textrm{i}/M_\odot<100.
\end{aligned}
\end{equation}

In our grid, stars with $M<1.3\,M_{\odot}$ possess a convective envelope already on the MS, and thus they skip the HG stage before entering the RGB. 

For stars with $M \le 2.0\,M_{\odot}$, the helium core remains degenerate on the RGB, and a helium flash occurs at the RGB tip. In this case, star allows the convective envelope to reach the core boundary and trigger the first dredge-up. Stars with $2.5\,M_{\odot} \le M \le 20\,M_{\odot}$ experience a short RGB phase with a non-degenerate core. In our grid, we skip the transitional models with $2.0\,M_{\odot}<M<2.5\,M_{\odot}$. As shown in \cite{PaperII}, for the same initial parameters, the critical transition model from degenerate to non-degenerate helium ignition occurs at $2.04\,M_{\odot}$.

The massive stars are usually defined as explodable $M>8\,M_{\odot}$ stars \citep{2006Natur.444..703C} or stars with $L>10^{4}\,L_{\odot}$ \citep{2026Univ...12..229H}. In our grid, stars with $M>20\,M_{\odot}$ start helium ignition before reaching the giant branch. In this work, for the convenience of conclusion of $E_\textrm{bind}$ distribution, we define the stars with $M>20\,M_{\odot}$ as massive stars in this work.

We classify stars in our grid as follows: $M<2.04\,M_{\odot}$ are low-mass stars, $2.04\,M_{\odot}<M<20\,M_{\odot}$ are intermediate-mass stars, and $M>20\,M_{\odot}$ are massive stars.

\subsection{CEE Energy Formalism}

\begin{figure*}[t!]
	\plotone{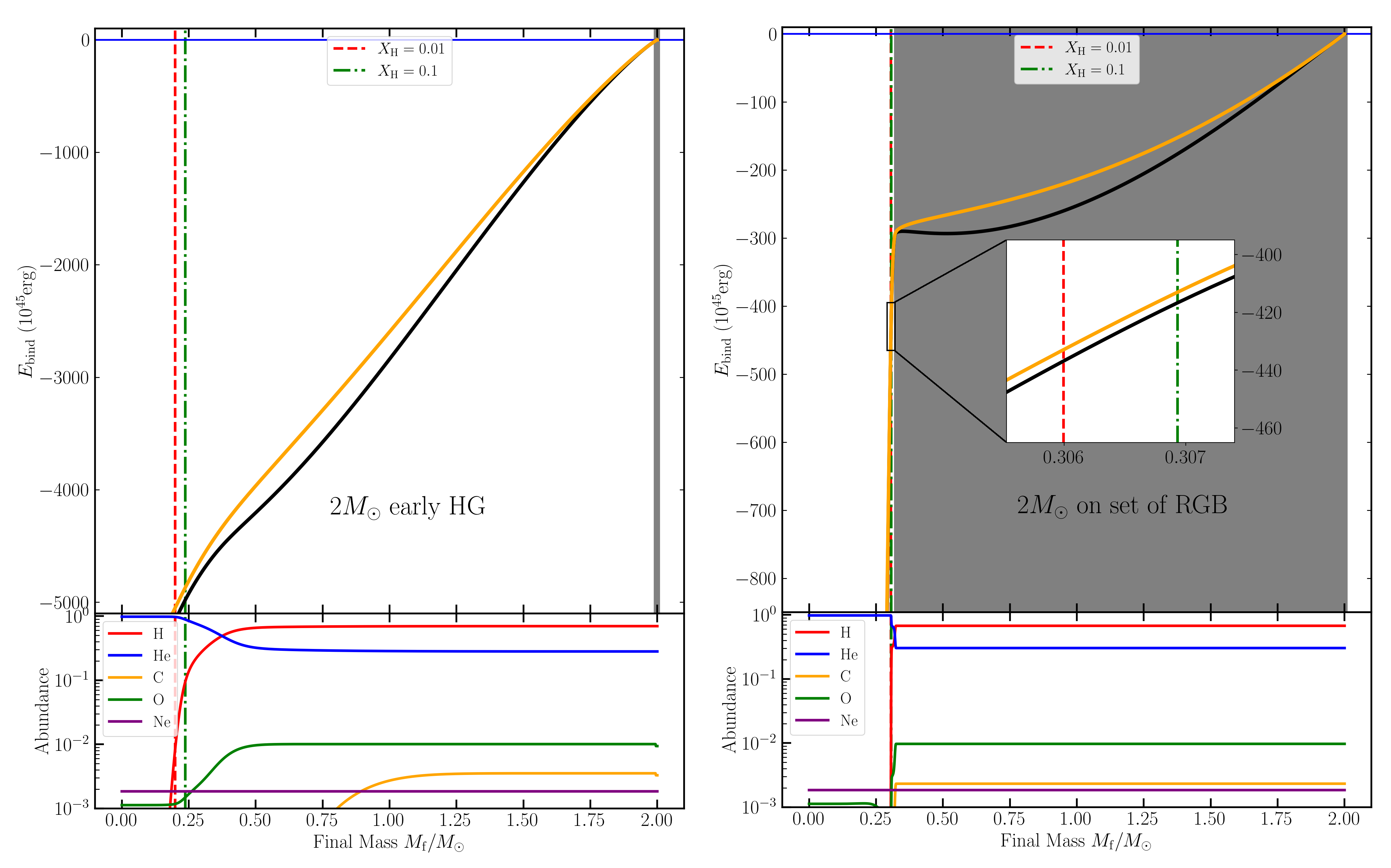}
	\caption{Binding energy of a $2.0\,M_{\odot}$ star as a function of the final mass $M_{\textrm{f}}$ after CEE. The left and right panels correspond to the early HG stage and the onset of the RGB, respectively. The solid black lines denote the results obtained from the initial CEE profile using Equation\,\ref{Ebind-i}, while the orange lines represent those from the adiabatic mass loss model using Equation\,\ref{Ebind-ad}. At the bottom of each panel, the elemental abundances are plotted on a logarithmic scale. Two definitions of the helium core boundary are adopted, corresponding to hydrogen mass fractions of $X_{\textrm{H}}=0.01$ and $X_{\textrm{H}}=0.1$, which are indicated by the red and green vertical lines, respectively.
		\label{Ebind-2.0}}
\end{figure*}

Surviving post-CEE binaries typically have very short orbital periods, often less than several days. This results from the ejection of the donor's envelope, which consumes orbital kinetic energy. The total orbital energy loss during CEE is given by
\begin{equation}
	\Delta E_\textrm{orb}= -\frac{GM_2}{2} \left( \frac{M_\textrm{f}}{a_\textrm{f}}-\frac{M_\textrm{i}}{a_\textrm{i}} \right), 
\end{equation}
where subscripts $\textrm{i}$ and $\textrm{f}$ denote the initial and final states before and after CEE, respectively. Here, $G$ is the gravitational constant, $a$ is the binary separation, $M$ is the donor mass, and $M_2$ is the accretor mass. In most cases, the increase in $M_2$ is negligible due to the short CEE timescale. The orbital energy $\Delta E_\textrm{orb}$ is used to overcome the envelope binding energy $E_\textrm{bind}$, and an efficiency parameter $\alpha_\textrm{CE}$ is introduced:
\begin{equation}
	E_\textrm{bind}=\alpha_\textrm{CE}\Delta E_\textrm{orb}. 
\end{equation}

A key issue is the definition of the envelope binding energy. One widely adopted model adopts the energy formalism that neglects structural changes and includes only gravitational potential and internal energy:
\begin{equation}\label{Ebind-i}
	E_\textrm{bind,i}=\int_{M_\textrm{f}}^{M_\textrm{i}} -\frac{GM_\textrm{i}\left ( r \right ) }{r} dm + \alpha_\textrm{th} \int_{M_\textrm{f}}^{M_\textrm{i}} u\,dm,
\end{equation}
where $-GM_\textrm{i}(r)/r$ and $u$ are the gravitational and internal energies per unit mass, evaluated at the donor's initial mass coordinate. The parameter $\alpha_\textrm{th}$ ($0 \le \alpha_\textrm{th} \le 1$) quantifies the fraction of internal energy contributing to envelope ejection. This formula uses the pre-CEE stellar profile, typically taken at the moment when the star fills its Roche lobe. In BPS studies, $M_\textrm{f}$ is often set to the core mass $M_\textrm{core}$, assuming complete envelope ejection in non-merger systems.

A subset of typical stellar models from the grid is selected for adiabatic mass loss calculations. These selected models have similar intervals in $\log R$ to cover the full evolutionary stages. The adiabatic mass loss model is a one-dimensional simulation that solves the dynamically unstable mass loss process. To mimic the CEE process, the model fixes the entropy profile in mass coordinates during rapid mass loss. 

During the mass loss in adiabatic models, the stripped envelope is removed adiabatically, while the remaining part of the star remains in hydrostatic equilibrium. As the envelope is stripped, the inner layers expand, whereas the heat remains constant owing to a fixed entropy profile. Other components of the internal energy, such as radiation energy, are directly determined by the specific physical parameters of the stellar structure. Consequently, compared with the initial pre-CEE profile, the absolute value of $E_\textrm{bind}$ decreases primarily for two reasons: the increase in radius $r$ of each mass shell, and the decrease in total stellar mass $M$. 

We follow method from \cite{PaperI} to compute the binding energy in the adiabatic mass loss model:
\begin{equation}\label{Ebind-ad}
	E_\textrm{bind,ad}=E_\textrm{total}\left ( M_\textrm{i} \right )-E_\textrm{total}\left ( M_\textrm{f} \right ) ,
\end{equation}
where the subscript $\textrm{ad}$ denotes the adiabatic mass loss model, and $E_\textrm{total}$ is the integrated binding energy from the surface to the centre:
\begin{equation}
	E_\textrm{total}\left ( M \right )=\int_{0}^{M} -\frac{GM\left ( r \right ) }{r} dm + \alpha_\textrm{th} \int_{0}^{M} u\,dm.
\end{equation}
Here, $E_\textrm{total}$ is evaluated for a specific stellar structure corresponding to an adiabatic model with a given amount of mass loss.

Figure\,\ref{Ebind-2.0} shows the $E_\textrm{bind}$ profile as a function of the final CEE mass $M_\textrm{f}$. The stars in both panels evolve from a $2.0\,M_{\odot}$ MS progenitor. The grey shaded area indicates the convective envelope. The black and orange solid lines represent $E_\textrm{bind,i}$ and $E_\textrm{bind,ad}$, computed from Equations\,\ref{Ebind-i} and\,\ref{Ebind-ad}, respectively. Owing to the adiabatic expansion effect, the orange lines are systematically higher than the black lines.

\begin{figure*}[t!]
	\plotone{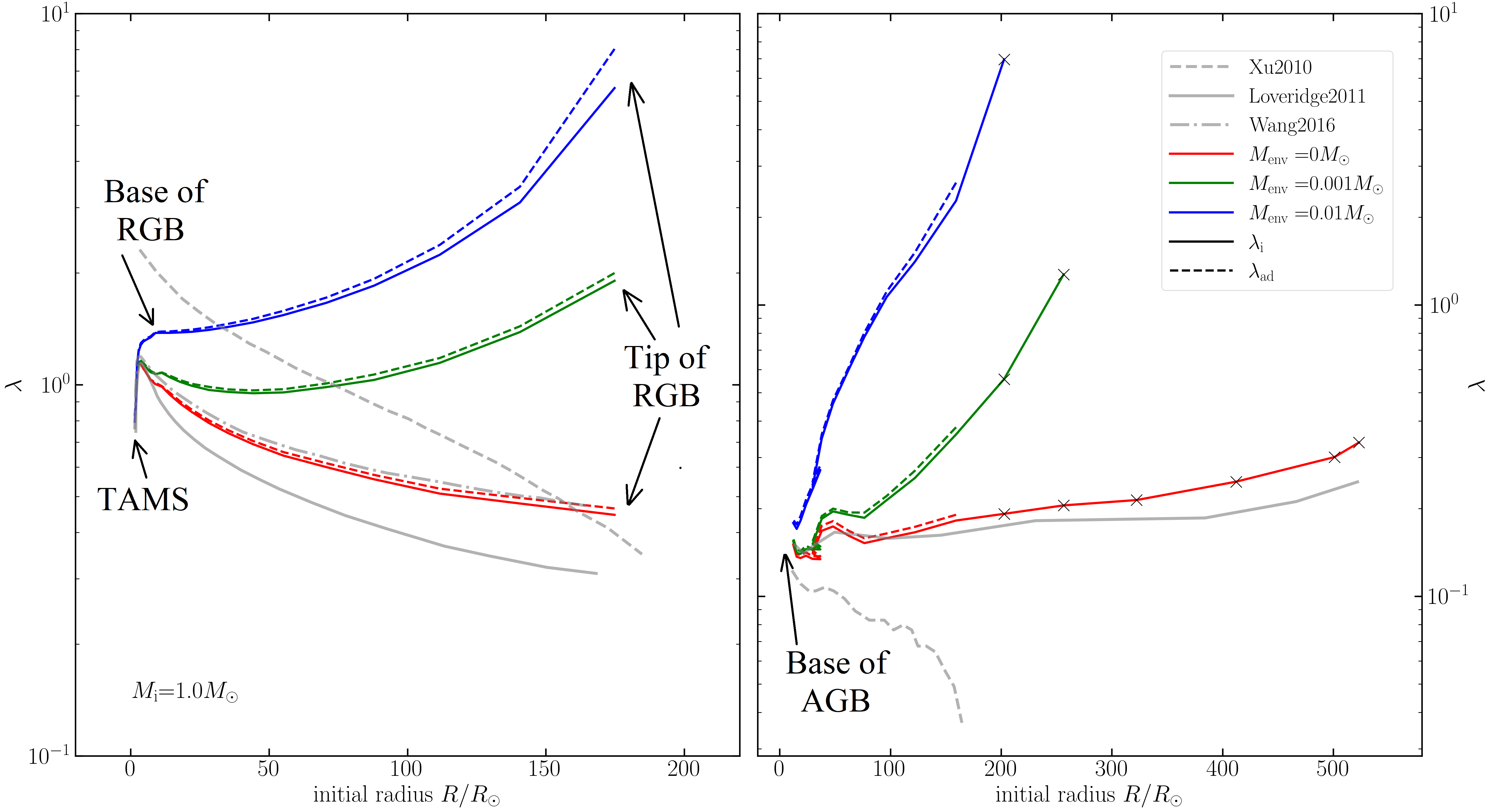}
	\caption{The $\lambda$ distribution for a $1.0\,M_{\odot}$ star as a function of the $R_\textrm{i}$ before CEE. The left panel is for the FGB stars, and the right one is for the AGB stars. The color of each line indicates the envelope residual mass $M_\textrm{env}$. The solid lines are derived from the initial CEE profiles, while the dashed lines are obtained from the adiabatic mass loss models. The grey lines represent $\lambda_\textrm{core}$ from previous studies, including \cite{2010ApJ...716..114X} (grey dashed lines), \cite{2011ApJ...743...49L} (solid lines), and \cite{2016RAA....16..126W} (dashed--dotted lines). Each line traces $\lambda$ from the ZAMS to the AGB, with key evolutionary stages marked in the figure. Black crosses highlight models with positive $E_\textrm{bind}$.
		\label{lamda_1M}}
\end{figure*}

In this work, we adopt two definitions of the core boundary, corresponding to hydrogen mass fractions of $X_\textrm{H}=0.01$ and $X_\textrm{H}=0.1$. The core mass $M_\textrm{core}$ is determined according to these two criteria. Both boundaries are marked in Figure\,\ref{Ebind-2.0} as red and green dashed lines. We note that the boundary positions vary at early evolutionary stages. After the star evolves on to the giant branch, $E_\textrm{bind}$ in the convective envelope becomes lower, and the gradient of $E_\textrm{bind}$ near the core boundary becomes steeper. We define the residual envelope mass after CEE as
\begin{equation}
    M_\textrm{env}=M_\textrm{f}-M_\textrm{core}.
\end{equation}

From the $E_\textrm{bind}$ profiles in Figure\,\ref{Ebind-2.0}, we find that both the envelope mass residual and the adiabatic expansion effect significantly affect $E_\textrm{bind}$ when the donor reaches the giant branch. In the following sections, we will quantify the detailed improvements arising from these two effects.

For convenience, $E_\textrm{bind}$ is often expressed in terms of a dimensionless parameter $\lambda$ \citep{1984nsf....8317916W,1990ApJ...358..189D}:
\begin{equation}\label{define_lam}
	E_\textrm{bind} \left ( M_\textrm{env} \right ) = -\frac{1}{\lambda \left ( M_\textrm{env} \right )} \cdot \frac{GM_\textrm{i}\left ( M_\textrm{i}-M_\textrm{f} \right ) }{ R_\textrm{i}},
\end{equation}
where $R_\textrm{i}$ is the initial radius before CEE, typically taken as the Roche-lobe radius at the onset of CEE. Figure\,\ref{lamda_1M} shows $\lambda \left ( M_\textrm{env} \right )$ for a $1\,M_\odot$ star. The left panel gives the results on the first giant branch (FGB, equivalent to HG plus RGB), and the right panel shows those for the AGB. Different colors represent different values of $M_\textrm{env}$. It is evident that $M_\textrm{env}$ has a significant impact on $\lambda \left ( M_\textrm{env} \right )$. Therefore, it is necessary to improve the determination of $\lambda$.

We define the total modification of $\lambda$ as
\begin{equation}\label{eq_F1_F2}
\begin{aligned}
	\lambda&=\lambda_\textrm{core} \cdot F_1 \cdot F_2, \\
    F_1&=\lambda_\textrm{i}\left ( M_\textrm{env} \right )/\lambda_\textrm{core},\\
    F_2&=\lambda_\textrm{ad}\left ( M_\textrm{env} \right )/\lambda_\textrm{i}\left ( M_\textrm{env} \right ),
\end{aligned}
\end{equation}
here $\lambda_\textrm{core}$ is the $\lambda$ value when $M_\textrm{env}=0\,M_\odot$, and $\lambda_\textrm{i}$ and $\lambda_\textrm{ad}$ are determined by the two prescriptions of $E_\textrm{bind}$ given in Equations\,\ref{Ebind-i} and \ref{Ebind-ad}, respectively. Through the above methods and our stellar evolution grid, we can obtain the improvement of $E_\textrm{bind}$ at different evolutionary stages. We present the results for Equation\,\ref{eq_F1_F2} in the next section.

\section{the improvement of $\lambda$} \label{sec:correction}

\begin{figure*}[htbp!]
	\centering
	\includegraphics[width=0.8\textwidth]{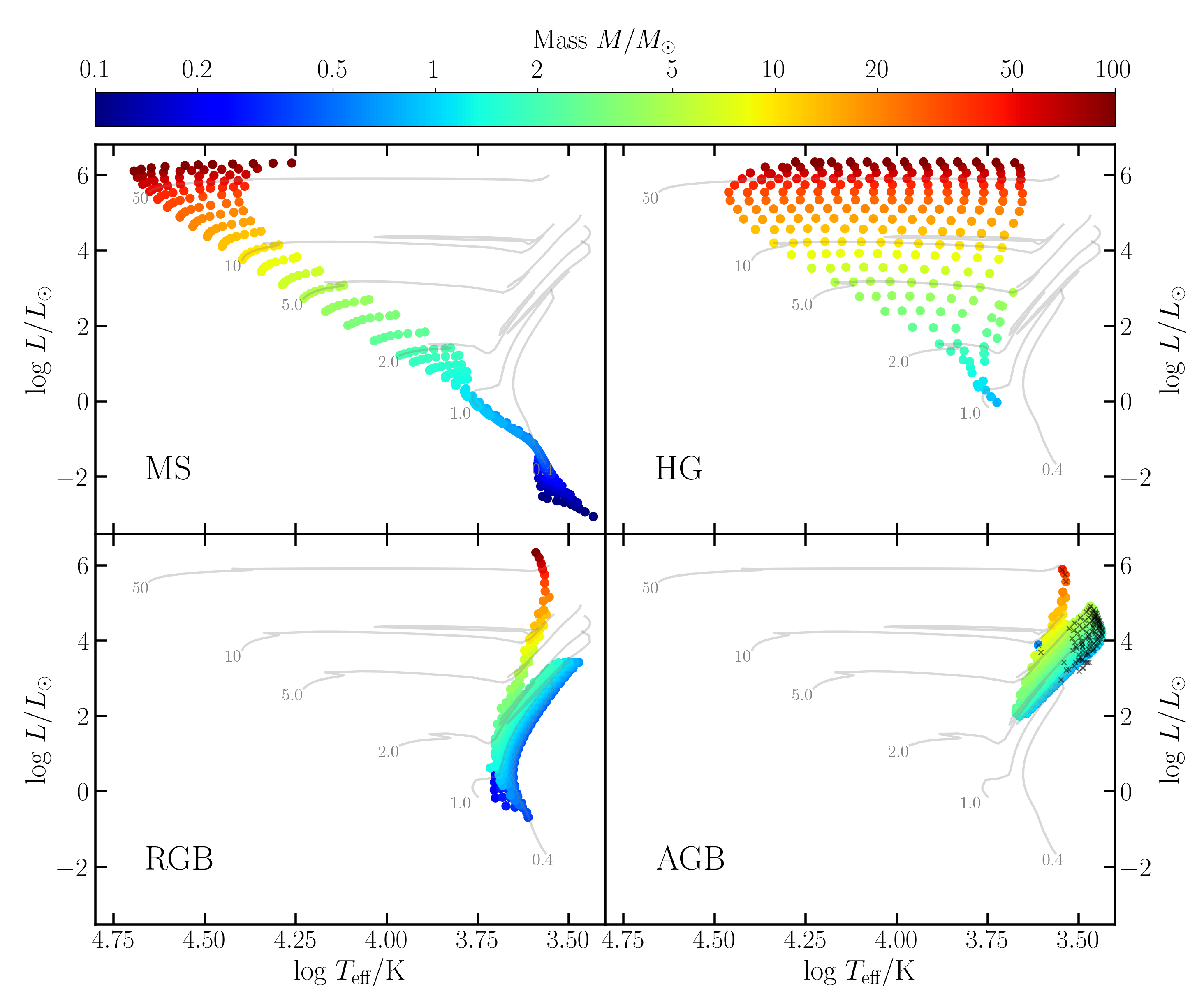}
	\caption{The full grid of adiabatic mass loss models on the HR diagram. The color scale represents the stellar mass on a logarithmic scale. The grid is divided into four evolutionary stages. The binding energy calculation covers the HG, RGB, and AGB stages. The black cross markers indicate models with positive $E_\textrm{bind}$ in the stellar envelope, which are excluded from the subsequent analysis. The grey solid lines show the evolutionary tracks of selected stars, with their masses labelled near the ZAMS.
		\label{HR_space}}
\end{figure*}

In the previous section, we presented the methods for computing $\lambda$. In this section, we use the adiabatic mass loss grid from \cite{PaperIII} to analyse the distributions of $\lambda_\textrm{core}$, $F_1$, and $F_2$. The full grid on the Hertzsprung--Russell (HR) diagram is shown in Figure\,\ref{HR_space}. For convenience in the subsequent analysis, we divide the models into MS, HG, RGB, and AGB stages. Models on the MS are excluded, as binary systems in this stage are generally believed to merge directly owing to the lack of a well-developed helium core. Our analysis therefore focuses on the HG, RGB, and AGB stages.

At certain stages near the tip of the giant branch, the convective envelope extends deeply into the stellar interior. Consequently, the envelope becomes very extended, and $E_\textrm{bind,i}$ becomes positive in some regions, which is similar to the results of \cite{1995MNRAS.274..964P}. These models, which are mainly late-AGB stars in our grid, are marked as black crosses in Figure\,\ref{HR_space}. We refer to these objects as \textit{super-thermal giants}. To avoid such unstable structures, we exclude all \textit{super-thermal giants} from the following analysis. We provide a fitting formula for the minimum radius of these \textit{super-thermal giants}:
\begin{equation}
    \log \left ( \frac{R_\textrm{sup}}{R_\odot} \right )  =\left [ \log \left ( \frac{M_\textrm{i}}{M_\odot} \right ) +0.444\right ] ^{0.435}+1.63,
\end{equation}
where the $R_\textrm{sup}$ is the minimum radius boundary of the \textit{super-thermal giants}. In our stellar evolution grid, these \textit{super-thermal giants} have masses in the range $0.5$ to $5\,M_{\odot}$, with a degenerate carbon--oxygen core.

\begin{table}[t]
    \centering
    \caption{Core boundary and envelope mass residual settings} 
    \label{tab:setting}
    \begin{tabular}{llll}
        \hline
{} & boundary & $M_\textrm{env}$ \\
        \hline
\textbf{Set-1} & $X_\textrm{H}=0.01$ & $\eta \cdot 1\,M_{\odot}$ \\
\textbf{Set-2} & $X_\textrm{H}=0.1$ & $\eta \cdot 1\,M_{\odot}$ \\
\textbf{Set-3} & $X_\textrm{H}=0.01$ & $\eta\cdot \,M_\textrm{i}$ \\
\textbf{Set-4} & $X_\textrm{H}=0.1$ & $\eta\cdot \,M_\textrm{i}$ \\
        \hline
    \end{tabular}
    \tablecomments{In this work, we set the envelope mass residual parameters $\eta \in \left \{ 0,  10^{-4}, 10^{-3}, 10^{-2}, 10^{-1} \right \} $.}
\end{table}

Based on the methods described in Section\,\ref{sec:method}, we adopt four parameter settings (listed in Table\,\ref{tab:setting}) to determine the $\lambda$ distribution. In the following subsections, we use these settings to compute $\lambda_\textrm{core}$, $F_1$, and $F_2$.

\begin{figure}[t!]
	\plotone{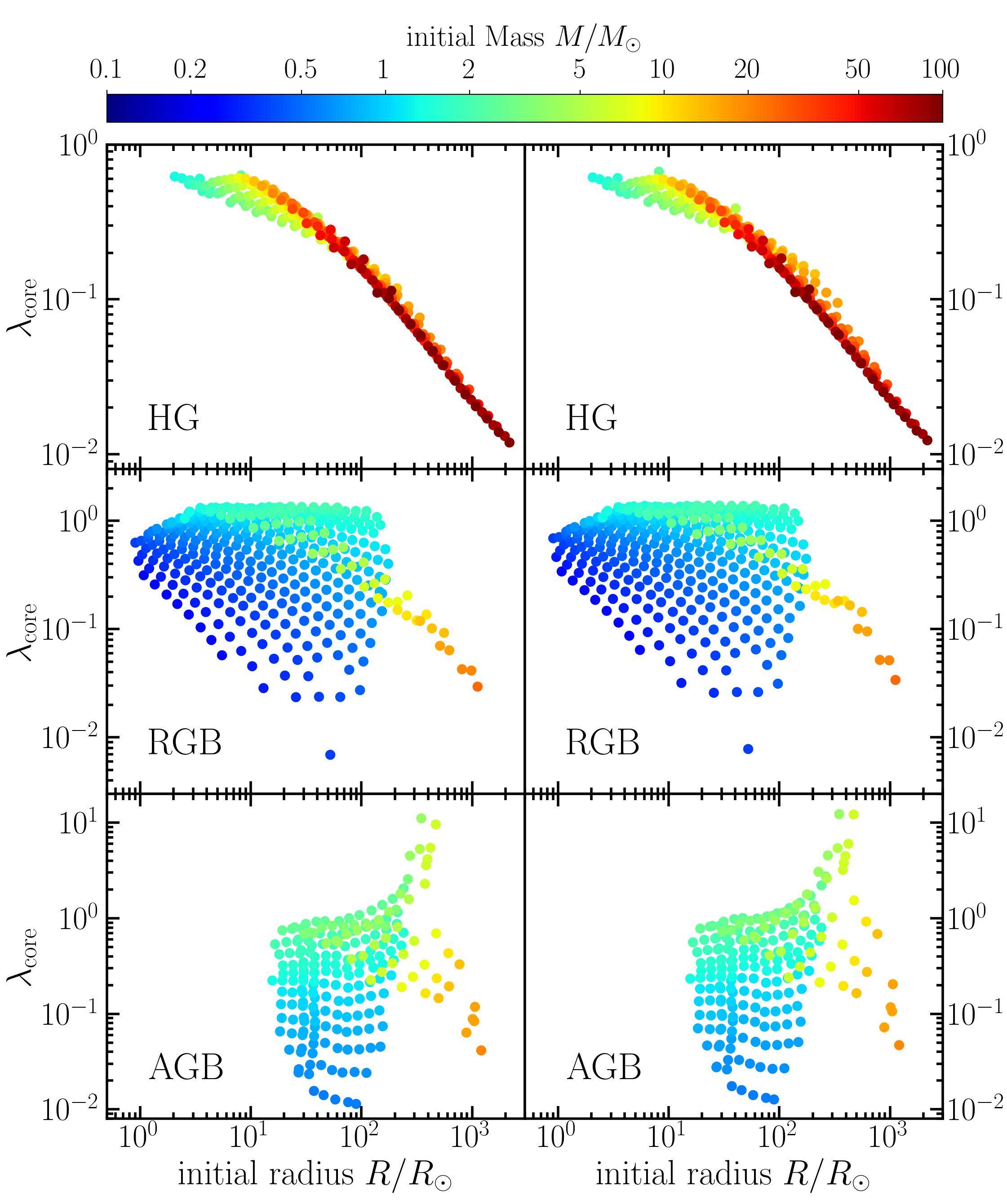}
	\caption{The $\lambda_\textrm{core}$ distribution for a $1.0\,M_{\odot}$ star as a function of the initial radius $R_\textrm{i}$ before CEE. The color scale represents the initial mass $M_\textrm{i}$. The left and right columns correspond to core boundaries defined by $X_\textrm{H}=0.01$ and $X_\textrm{H}=0.1$, respectively. The evolutionary stage before CEE is indicated in each subplots. The relative error of most of $\lambda_\textrm{core}$ in this two boundaries is within 15\%.
		\label{lam_core}}
\end{figure}

\begin{figure*}[htbp!]
	\centering
	\includegraphics[width=0.745\textwidth]{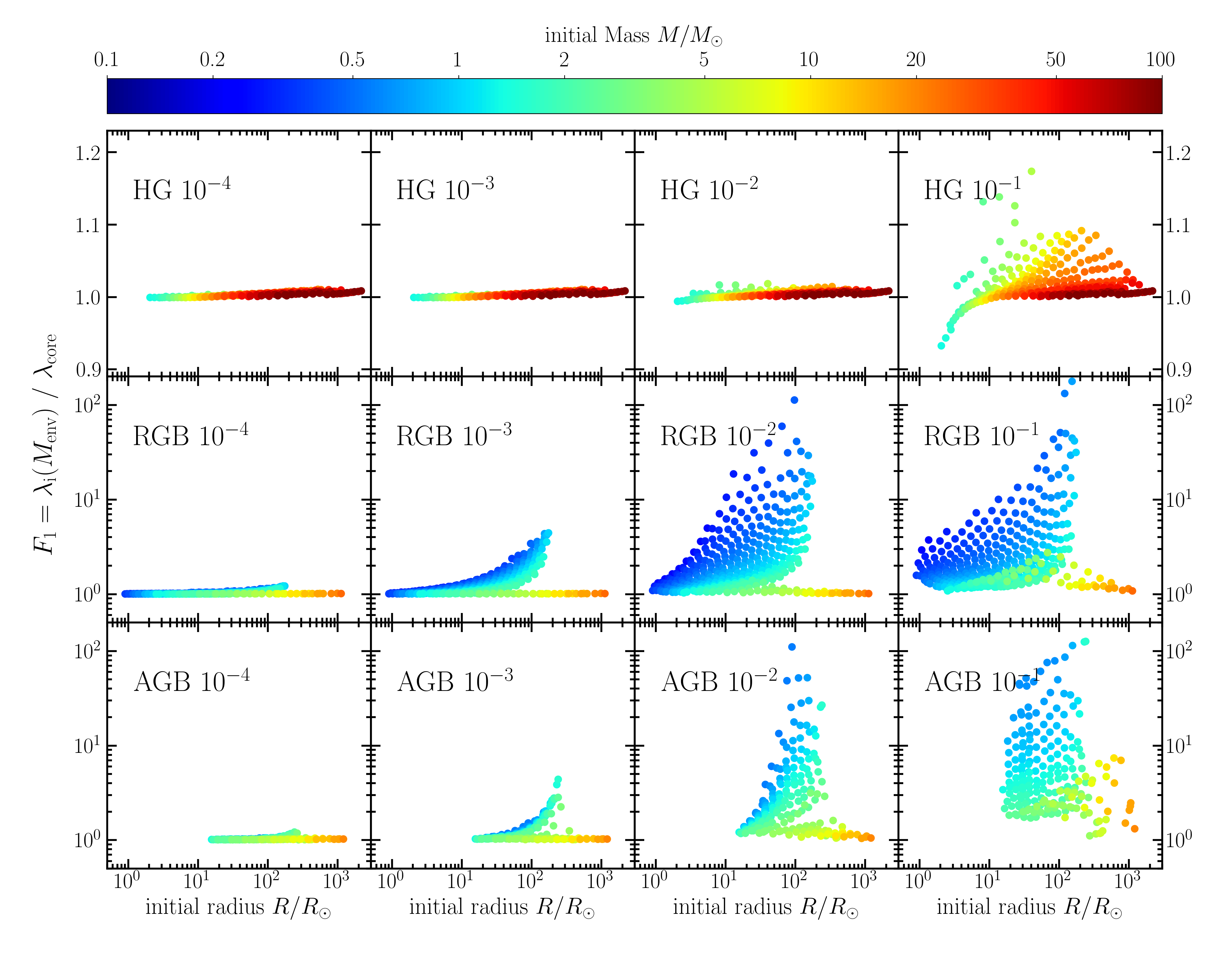}
	\caption{The $F_1$ distribution obtained using \textbf{Set-1} as a function of the initial radius $R_\textrm{i}$ before CEE. The color scale represents the initial mass $M_\textrm{i}$. Each column corresponds to the same CEE envelope residual $\eta$ defined in Table\,\ref{tab:setting}.
		\label{F1_Absolute}}
\end{figure*}

\vspace{1em}

\begin{figure*}[htbp!]
    \centering
	\includegraphics[width=0.745\textwidth]{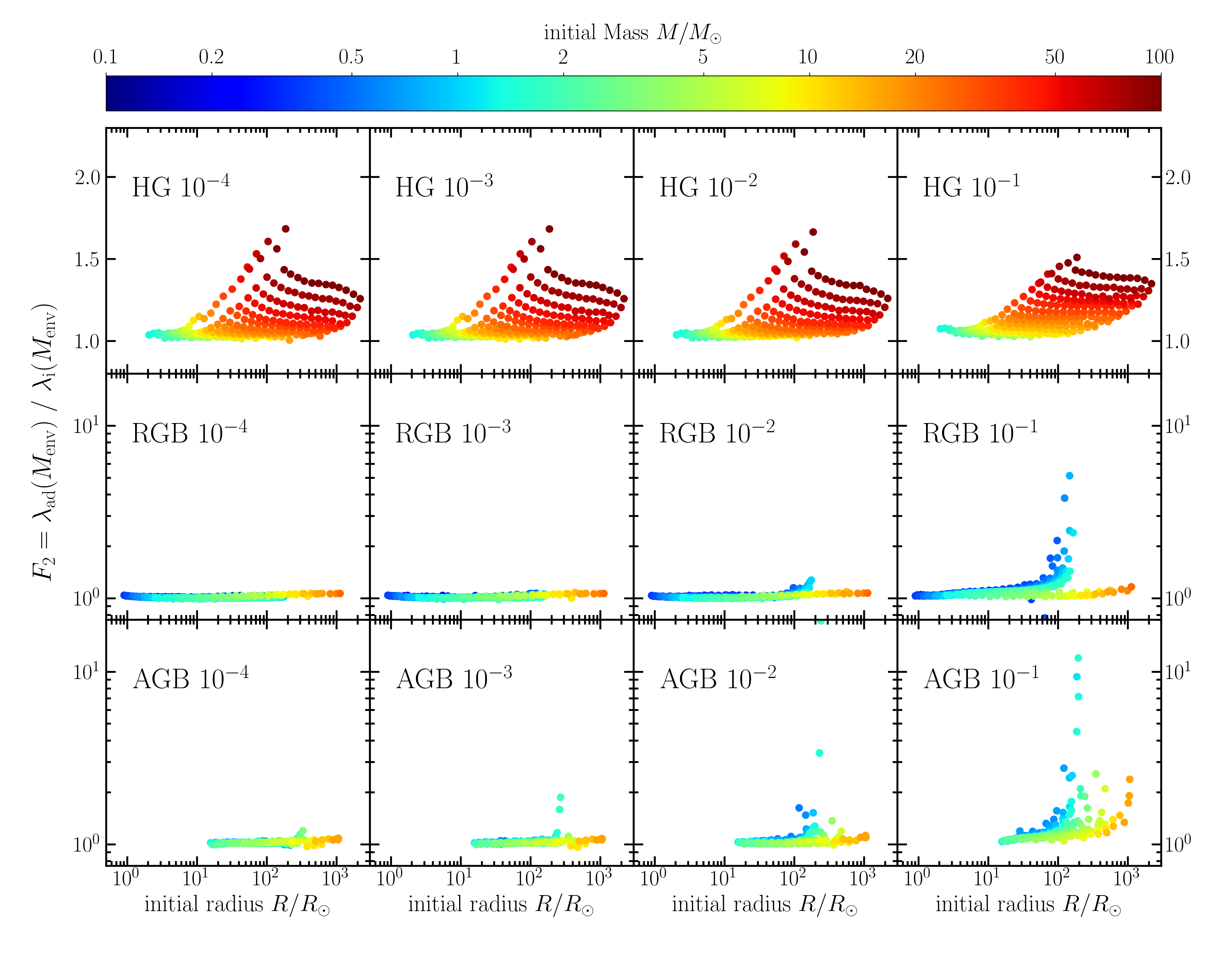}
	\caption{The $F_2$ distribution obtained using \textbf{Set-3} as a function of the initial radius $R_\textrm{i}$ before CEE. The color scale represents the initial mass $M_\textrm{i}$. Each column corresponds to the same CEE envelope residual $\eta$ defined in Table\,\ref{tab:setting}.
		\label{F2_Relative}}
\end{figure*}

\subsection{$\lambda_\textrm{core}$ Distribution} \label{sec:lam_core}

Based on Equations\,\ref{Ebind-i} and\,\ref{define_lam}, $\lambda_\textrm{core}$ is determined by setting $M_\textrm{f}=M_\textrm{core}$ (i.e. $M_\textrm{env}=0\,M_\odot$). In Figure\,\ref{lamda_1M}, the red solid lines show $\lambda_\textrm{core}$ for a $1.0\,M_{\odot}$ star on the RGB and AGB stages.

Our results are in good agreement with previous studies, such as \cite{2011ApJ...743...49L} and \cite{2016RAA....16..126W}. Those works included the effects of various stellar wind prescriptions, which are not considered in this study. Nevertheless, the $\lambda_\textrm{core}$ distributions are quite similar, particularly for low-mass stars. Compared with the pioneering work of \cite{2010ApJ...716..114X}, our results yield lower $\lambda_\textrm{core}$ values at early evolutionary stages, implying a larger absolute value of $E_\textrm{bind}$. This difference mainly arises from the full ionization of several metals, such as carbon and nitrogen, in their calculations.

Figure\,\ref{lam_core} summarises the $\lambda_\textrm{core}$ distributions at different evolutionary stages. For low-mass stars, $\lambda_\textrm{core}$ decreases as the star evolves. This is because low-mass RGB stars evolve on the nuclear timescale owing to the critical core mass required for the helium flash at the tip of RGB, during which the core mass increases significantly. For the other stages of intermediate- and massive RGB/AGB stars, which evolve on the thermal timescale, the increase in core mass is small. In this case, the decrease in gravitational potential energy caused by envelope expansion dominates, and $\lambda_\textrm{core}$ increases as the star evolves.

In Figure\,\ref{lam_core}, the left and right columns correspond to core boundaries defined by $X_\textrm{H}=0.01$ and $X_\textrm{H}=0.1$, respectively. The difference between the two sets of $\lambda_\textrm{core}$ results is not significant. On the RGB and AGB, the gradient of $X_\textrm{H}$ is sufficiently steep that $M_\textrm{core}$ is nearly identical under both definitions. From our full grid calculations, the difference in $\lambda_\textrm{core}$ between the two definitions is less than 15\% for most stars. The largest discrepancy occurs for intermediate-mass RGB and early-AGB stars, before the first dredge-up.

\subsection{Envelope Residual Improvement of $F_1$} \label{sec:f1}

In this subsection, we discuss the effect of the envelope residual at the end of CEE. We still adopt the classical formalism (Equation\,\ref{Ebind-i}) to compute $E_\textrm{bind}$ from the initial pre-CEE profile. The factor $F_1$ is defined in Equation\,\ref{eq_F1_F2}.

As Figure\,\ref{lamda_1M} indicates, the envelope residual reverses the decreasing trend of low-mass RGB stars, owing to the rapidly increasing density gradient near the core boundary as the envelope expands along the RGB. The absolute value of $E_\textrm{bind}$ decreases as both $M_\textrm{env}$ and $R_\textrm{i}$ increase.

Figure\,\ref{F1_Absolute} shows the distribution of $F_1$ (using \textbf{Set-1}) at different evolutionary stages. The stellar models in each stage correspond to those shown in Figure\,\ref{HR_space}. In each column, the same value of $\eta$ (given in the upper-left corner) is used to determine the CEE envelope residual $M_\textrm{env}$.

As seen in Figure\,\ref{F1_Absolute}, $F_1$ becomes significant at later evolutionary stages, particularly for models with a deep convective envelope, such as RGB-tip stars (for example, a $1\,M_\odot$ on tip-RGB with $M_\textrm{env}=0.01\,M_\odot$, $F_1\sim 15$). On the HG stage, $F_1$ remains small, typically below 1.1. As $M_\textrm{env}$ increases, some RGB and AGB stars reach $F_1>100$. This implies that the corresponding $E_\textrm{bind}$, including the envelope residual, can be as low as 1/100 of the value previously expected, suggesting that the CE would be much easier to eject.

\subsection{Adiabatic Correction of $F_2$} \label{sec:f2}

In this subsection, we further discuss the adiabatic expansion effect. We adopt adiabatic mass loss models to predict the correction to the binding energy arising from adiabatic expansion during CEE. For each initial pre-CEE model, mass is removed under adiabatic conditions. This approach effectively simulates the structural evolution of the remaining stellar envelope during the CEE process. We use the adiabatic formalism (Equation\,\ref{Ebind-ad}) to compute $E_\textrm{bind}$ from the adiabatic models. The factor $F_2$ is defined in Equation\,\ref{eq_F1_F2}.

Following the same format as Figure\,\ref{F1_Absolute}, Figure\,\ref{F2_Relative} shows the distribution of $F_2$ (using \textbf{Set-3}) for the same sequence of $\eta$ values. At most evolutionary stages, the $F_2$ effect is not significant. For massive stars, adiabatic expansion yields $F_2<1.75$ at the early HG stage. The most pronounced effect of $F_2$ occurs near the RGB tip with thick a envelope residual, where $F_2$ rapidly approaches 10, owing to the high entropy of the extended convective envelope. The AGB branch exhibits a similar trend. Here we note that $F_2$ shows a rapid increase when $M_\textrm{env} \gtrsim 0.01\,M_\odot$ for most giant-branch stars, which is mainly due to the entropy profile reaching its maximum value near this region. 

Since our grid stops before the models reach the \textit{super-thermal giant} phase, the $F_2$ behaviour at the late TP-AGB stage remains unexplored and may require special treatment for CEE with TP-AGB donors.

As the $F_2$ results indicate, adiabatic expansion has little effect on $E_\textrm{bind}$ for most evolutionary stages (for example, a $1\,M_\odot$ on tip-RGB with $M_\textrm{env}=0.01\,M_\odot$, $F_2\sim 1.5$). However, near the RGB tip and the late AGB, if the envelope residual satisfies $\eta>0.01$, adiabatic expansion can still provide $F_2>2$.

\section{Interpolations and fitting formulae on each stage} \label{sec:fit}

In the preceding sections, we have presented detailed calculations of $\lambda_\textrm{core}$, $F_1$, and $F_2$. In this section, we provide interpolation routines and fitting formulae for these results, which are useful for BPS and other binary studies. To ensure accuracy and reliability, the interpolations are performed separately for different evolutionary stages. We first apply radial basis function (RBF) interpolation to resample the original data from Section\,\ref{sec:lam_core}, and then adopt polynomial functions to fit the formulae for each resampled table.

The definitions of the evolutionary stage boundaries have been given in Section\,\ref{sec:method}. Here, we place models with $M_\textrm{i}>20\,M_{\odot}$ into a separate channel for massive stars. These stars ignite helium already at the early HG stage. In other studies that include stellar wind effects, such massive stars are likely to lose their hydrogen envelopes and become Wolf--Rayet stars. Accordingly, we divide the parameter space into four channels: HG, RGB, AGB, and massive stars.

\subsection{RBF Interpolation}\label{subsec:RBF}
To ensure that the grid accounts for variations in stellar models arising from different initial parameters, the interpolated grid should be smooth, avoid overfitting, and possess reasonable extrapolation behaviour. In this work, we use three-dimensional radial basis function (RBF) interpolation \citep{Fasshauer2007,Wahba1990} from \href{https://docs.scipy.org/doc/scipy/reference/generated/scipy.interpolate.Rbf.html}{\textbf{Scipy}} to compute the results over the parameter space $\left ( \left [ M_\textrm{i} \right ], \left [ R_\textrm{i} \right ], \left [ \eta \right ] \right )$, where $\left [ M_\textrm{i} \right ]\equiv \log M_\textrm{i}/M_\odot$, $\left [ R_\textrm{i} \right ]\equiv \log R_\textrm{i}/R_\odot$, and $\left [ \eta \right ]\equiv \log \eta$ (similar with Equation\,\ref{eq_BRGB}, following the definitions in Table\,\ref{tab:setting}). The resampled parameter space also facilitates function fitting in the next step. After testing, the interpolation performs well with the radial basis function kernel='cubic' and smooth=0.1. We have uploaded the interpolation tables and the corresponding codes for each set listed in Table\,\ref{tab:setting} to the \href{https://github.com/CygX-3-stargate0/Adiabatic_CEE_solver.git}{\textbf{website}} on GitHub. This code allows users to obtain $\lambda_\textrm{core}$, $F_1$, and $F_2$ for CEE systems within the range $-4\le \log M_\textrm{env}/M_\odot\le -1$.

\begin{table*}[htbp]
  \centering
  \caption{The accuracy of RBF interpolation.}
  \label{tab:RBF_acc}
  \begin{center}
    \begin{tabular}{lcrrrrrrrr}
      \hline
      &  &  \multicolumn{2}{c}{$\lambda_\textrm{core}$} & \multicolumn{2}{c}{$F_1$} & \multicolumn{2}{c}{$F_2$} & \multicolumn{2}{c}{$\lambda_\textrm{core}\cdot F_1 \cdot F_2$} \\
      &  & $\Delta _{0.1}$ & $\Delta _{0.2}$ & $\Delta _{0.1}$ & $\Delta _{0.2}$ & $\Delta _{0.1}$ & $\Delta _{0.2}$ & $\Delta _{0.1}$ & $\Delta _{0.2}$ \\
      \hline
      \multirow{4}{*}{\textbf{Set-1}} & HG & 94.49\% & 98.73\% & 100.00\% & 100.00\% & 99.58\% & 100.00\% & 94.92\% & 97.03\% \\
       & RGB & 91.78\% & 98.69\% & 93.58\% & 97.17\% & 97.38\% & 98.14\% & 85.91\% & 94.89\% \\
       & AGB & 50.00\% & 72.60\% & 66.87\% & 76.58\% & 76.38\% & 83.74\% & 45.71\% & 65.95\%\\
       & massive & 94.59\% & 100.00\% & 100.00\% & 100.00\% & 99.32\% & 100.00\% & 95.10\% & 97.13\% \\
       \hline
      \multirow{4}{*}{\textbf{Set-2}} & HG & 93.64\% & 98.73\% & 100.00\% & 100.00\% & 100.00\% & 100.00\% & 92.80\% & 97.46\% \\
       & RGB & 93.02\% & 98.41\% & 93.65\% & 97.31\% & 97.38\% & 98.14\% & 88.19\% & 95.17\% \\
       & AGB & 48.73\% & 73.35\% & 67.07\% & 77.10\% & 75.99\% & 83.18\% & 45.80\% & 64.94\% \\
       & massive & 94.59\% & 99.32\% & 100.00\% & 100.00\% & 100.00\% & 100.00\% & 95.10\% & 97.30\% \\
       \hline
      \multirow{4}{*}{\textbf{Set-3}} & HG & 94.49\% & 98.73\% & 99.68\% & 100.00\% & 99.68\% & 100.00\% & 93.64\% & 98.73\% \\
       & RGB & 91.33\% & 98.44\% & 92.21\% & 96.68\% & 97.49\% & 98.24\% & 85.65\% & 94.85\% \\
       & AGB & 50.10\% & 72.70\% & 65.85\% & 78.43\% & 77.20\% & 82.92\% & 46.32\% & 66.05\% \\
       & massive & 94.59\% & 100.00\% & 97.47\% & 98.31\% & 99.49\% & 100.00\% & 91.39\% & 96.45\% \\
       \hline
      \multirow{4}{*}{\textbf{Set-4}} & HG & 93.64\% & 98.73\% & 99.68\% & 100.00\% & 100.00\% & 100.00\% & 91.95\% & 97.67\% \\
       & RGB & 92.62\% & 98.17\% & 92.62\% & 96.75\% & 97.36\% & 98.24\% & 87.47\% & 95.13\% \\
       & AGB & 48.83\% & 73.45\% & 65.55\% & 77.51\% & 76.80\% & 82.78\% & 45.59\% & 65.86\% \\
       & massive & 94.59\% & 99.32\% & 97.47\% & 98.14\% & 100.00\% & 100.00\% & 91.22\% & 96.45\% \\
      \hline
    \end{tabular}
    \end{center}
\end{table*}

Using the RBF interpolation tables, we then interpolate the raw data from Section\,\ref{sec:correction} and record the points that satisfy the following criterion:
\begin{equation}
    \left | \frac{ D_\textrm{predict}-D_\textrm{raw}}{D_\textrm{raw}}  \right | < f,\quad D\in \left \{ \lambda_\textrm{core} , F_1 , F_2 \right \}.
\end{equation}
We define $\Delta _f$ as the fraction of samples satisfying this criterion relative to the total number of predicted samples. We adopt $\Delta _{0.1}$ and $\Delta _{0.2}$ to evaluate the interpolation accuracy in Table\,\ref{tab:RBF_acc}. According to our error analysis, over 90\% of the interpolated data have errors within $f=0.1$. The main issue occurs on the AGB branch, where a peak in $E_\textrm{bind}$ near the base of the convective envelope leads to a non-monotonic increase in $\lambda$ with increasing $M_\textrm{env}$. We provide a detailed discussion of this issue in Appendix\,\ref{apd.agb}.

\subsection{Polynomial Function}\label{subsec:poly}

\begin{table}[htbp]
  \centering
  \caption{Accuracy of the 3D-polynomial fitting formula and the 3D-RBF interpolation in a selected parameter space.}
  \label{tab:poly_acc}
    \begin{tabular}{rccccc}
      \hline
      \textbf{Set-3} &  & HG & RGB & AGB & massive \\
      \hline
       & $M_\textrm{i}/M_\odot$ & [1,100] & [0.8,10] & [0.8,10] & [8,100] \\
       & $R_\textrm{i}/R_\odot$ & [2,2000] & [1,400] & [15,200] & [10,2000] \\
       & $\eta$ & \multicolumn{4}{c}{[0.0001,0.1]} \\
       \hline
      \multirow{2}{*}{Poly} & $\Delta_{0.1}$ & 87.61\% & 76.27\% & 58.55\% & 91.34\% \\
       & $\Delta _{0.2}$ & 96.62\% & 92.49\% & 87.48\% & 96.65\% \\
       \hline
      \multirow{2}{*}{RBF} & $\Delta_{0.1}$ & 94.03\% & 89.34\% & 64.02\% & 91.73\% \\
       & $\Delta _{0.2}$ & 97.75\% & 97.30\% & 88.18\% & 96.46\% \\
      \hline
    \end{tabular}
\end{table}

Although the interpolation tables have been provided in the previous section (Section\,\ref{subsec:RBF}), their direct use in some BPS codes, such as BSE, remains difficult and may consume substantial computational resources. In this subsection, we therefore provide a three-dimensional polynomial fitting formula for $\lambda_\textrm{ad}=\lambda_\textrm{core}\cdot F_1\cdot F_2$, based on the interpolation tables from \textbf{Set-3}. The polynomial function takes the following form:
\begin{equation}
    \lambda_\textrm{ad}=\sum_{i+j+k \le N}^{} C_{i,j,k}\left [ M_\textrm{i} \right ]^i \left [ R_\textrm{i} \right ]^j \left [ \eta \right ]^k .
\end{equation}
where $N$ is the degree of the polynomial. The coefficient tables $C_{i,j,k}$ for each evolutionary stage are also uploaded to our \href{https://github.com/CygX-3-stargate0/Adiabatic_CEE_solver.git}{\textbf{website}}. Although the above formulae are relatively complex, they still provide fast computation in BPS codes. To ensure accuracy, we have deliberately reduced the coverage range of the parameter space, as summarised in Table\,\ref{tab:poly_acc}. We have slightly adjusted the mass ranges for different types of stars to reduce the error.

In addition, we also provide two-dimensional fitting formulae for fixed $\eta$ values for user convenience. For example, the following is the $\lambda_\textrm{ad}$ fitting formula for RGB stars at $\eta=0.01$:
\begin{equation}
\begin{aligned}
&\log \lambda_\textrm{ad} = \\
&0.027 + 0.083\left [ R_\textrm{i} \right ]+ 0.159\left [ M_\textrm{i} \right ]+ 0.404\left [ R_\textrm{i} \right ]^2 \\
&- 0.347\left [ M_\textrm{i} \right ]\left [ R_\textrm{i} \right ]- 0.409\left [ M_\textrm{i} \right ]^2- 0.548\left [ R_\textrm{i} \right ]^3 \\
&+ 1.196\left [ M_\textrm{i} \right ]\left [ R_\textrm{i} \right ]^2- 1.815\left [ M_\textrm{i} \right ]^2\left [ R_\textrm{i} \right ]- 0.607\left [ M_\textrm{i} \right ]^3\\
&+ 0.185\left [ R_\textrm{i} \right ]^4- 0.426\left [ M_\textrm{i} \right ]\left [ R_\textrm{i} \right ]^3+ 0.273\left [ M_\textrm{i} \right ]^2\left [ R_\textrm{i} \right ]^2\\
&+ 0.397\left [ M_\textrm{i} \right ]^3\left [ R_\textrm{i} \right ]
+ 0.634\left [ M_\textrm{i} \right ]^4.
\end{aligned}
\end{equation}

\begin{figure*}[t]
	\plotone{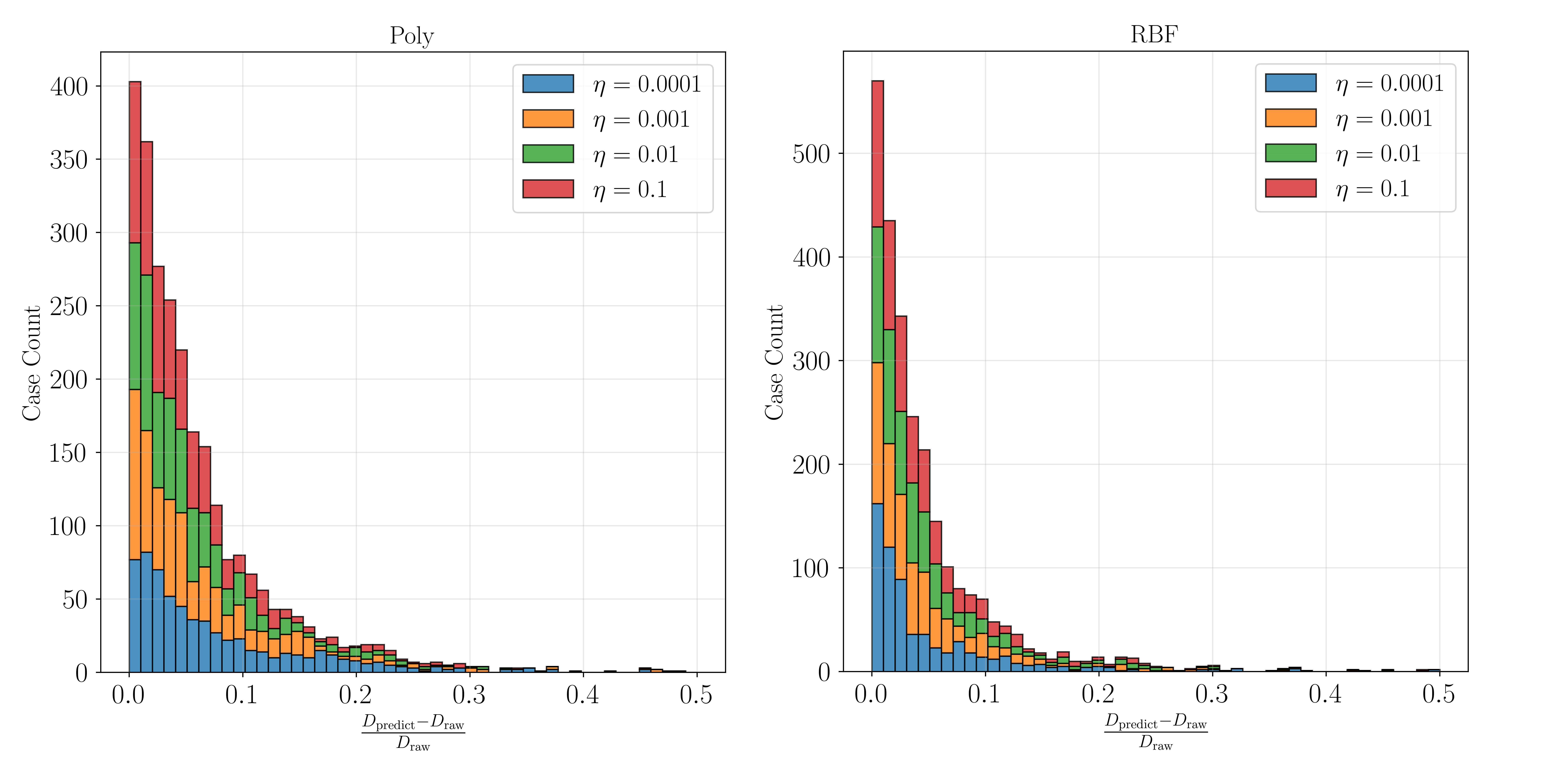}
	\caption{The error distributions of the polynomial fitting and the RBF interpolation relative to the raw data. The data source is the same as Table\,\ref{tab:poly_acc}. The colors represent the different envelope residuals.
		\label{err_Set3}}
\end{figure*}

Table\,\ref{tab:poly_acc} compares the accuracy of the RBF interpolation and the polynomial fitting formulae relative to the original data, within the reduced parameter space covered by the polynomial fits. To achieve better accuracy and reduce the value of $N$, we have adjusted the mass and radius ranges for different stages, as summarised in Table\,\ref{tab:poly_acc}. Figure\,\ref{err_Set3} shows the error distributions of the polynomial fitting and the RBF interpolation relative to the raw data. The colors represent different envelope residuals. The error data are the same as those in Table\,\ref{tab:poly_acc}, but the evolutionary stages are not distinguished. 
These results confirm that the fitting formulae are in good agreement with both the RBF interpolation and the raw data from Section\,\ref{sec:correction}.

\section{discussion}\label{discuss}

In the preceding sections, we have introduced the effects of the CEE envelope mass residual and adiabatic expansion. Our results show that even a very thin envelope residual ($M_\textrm{env}$) can cause significant deviations in the stellar binding energy $E_\textrm{bind}$. We have also provided interpolation grids and fitting formulae for different $M_\textrm{env}$ values, which accurately cover the $\lambda$ parameter for CEE with HG, RGB, and massive donors.

We have compared the $\lambda_\textrm{core}$ values in detail. When adopting different helium core boundaries $X_\textrm{H}=0.01$ or $0.1$, the resulting $\lambda_\textrm{core}$ shows little difference. On the HG and massive-star stages, the donor maintains a radiative envelope, leading to a nearly linear variation of $E_\textrm{bind}$ with $M_\textrm{f}$ (see Figure\,\ref{Ebind-2.0}). On the RGB or AGB, mixing at the convective boundary makes the core mass nearly identical under the two definitions. Our full grid calculations indicate that the difference in $\lambda_\textrm{core}$ between the two core definitions is within 15\% for most stars.

It has long been recognised that CEE may leave residual envelopes. For example, sdB stars are composed primarily of helium, yet most sdB spectra show hydrogen-rich features \citep{2016PASP..128h2001H}. Asteroseismic studies have provided direct observational evidence for the presence of $M_\textrm{env}$ after CEE \citep{2012A&A...539A..12F}. Typical systems include hot subdwarf stars and low-mass white dwarfs. Observations indicate that the surfaces of these stars are covered by hydrogen envelopes with masses in the range $10^{-6}$ to $10^{-2}\,M_{\odot}$. It is generally believed that the CEE channel can efficiently produce systems with such thin hydrogen envelopes. Meanwhile, 3D hydrodynamical simulations agree that CEE can effectively eject the envelope, but the envelope residuals are strongly influenced by the stellar equation of state \citep{2020A&A...644A..60S,2020A&A...642A..97K,2022MNRAS.512.5462L}.

It should be noted, however, that the observed hydrogen-envelope mass may differ significantly from the actual envelope mass at the end of CEE. Post-CEE hydrogen burning on the helium-core surface may further reduce the envelope thickness. Under this assumption, white dwarfs could burn their envelopes more rapidly owing to their higher temperatures at the helium-core boundary. Moreover, the stellar radius and surface gravity are sensitive to the hydrogen-envelope thickness, which can affect the generation of strong stellar winds. Even if a post-CEE star retains a thick envelope, subsequent hydrogen burning and winds may rapidly remove it, leaving a thinner envelope. Therefore, observational estimates of the hydrogen-envelope mass may only provide a lower limit to the true residual mass at the end of CEE. Finally, mixing processes at the base of the envelope may also contribute to a reduction in the hydrogen-envelope thickness \citep{2024RAA....24e5003J}.

If the donor indeed retains a significant envelope at the end of CEE, then $E_\textrm{bind}$ should be much lower than previously estimated. According to our results, the envelope residual may affect the properties of many PCEB systems: CE mergers become less likely, and the orbital-period distribution of PCEBs should shift towards longer periods.

In the standard CE evolution theory, the final orbital-period distribution of PCEBs depends on $\alpha_\textrm{CE}\lambda$. Currently, the formation of some PCEB systems requires different $\alpha_\textrm{CE}$ values. For short-period WD binaries, $\alpha_\textrm{CE} \sim 1/3$ \citep{2022MNRAS.513.3587Z,2023MNRAS.518.3966S}. For long-period oxygen--neon WD binaries, $\alpha_\textrm{CE} \sim 1$ \citep{2024A&A...686A..61B}. For massive stars, black-hole X-ray binaries suggest $\alpha_\textrm{CE} \gtrsim 4$ \citep{2026ApJ..1004...31L}. \cite{2024ApJ...961..202G} suggested that $\alpha_\textrm{CE}$ need not be constant and may vary as a function of the initial mass ratio. This suggests that $E_\textrm{orb}$ near the core may be inappropriate. The envelope residual improvement presented in this work may help to find the correct $\alpha_\textrm{CE}$ in BPS simulations.

\begin{figure}[t!]
	\plotone{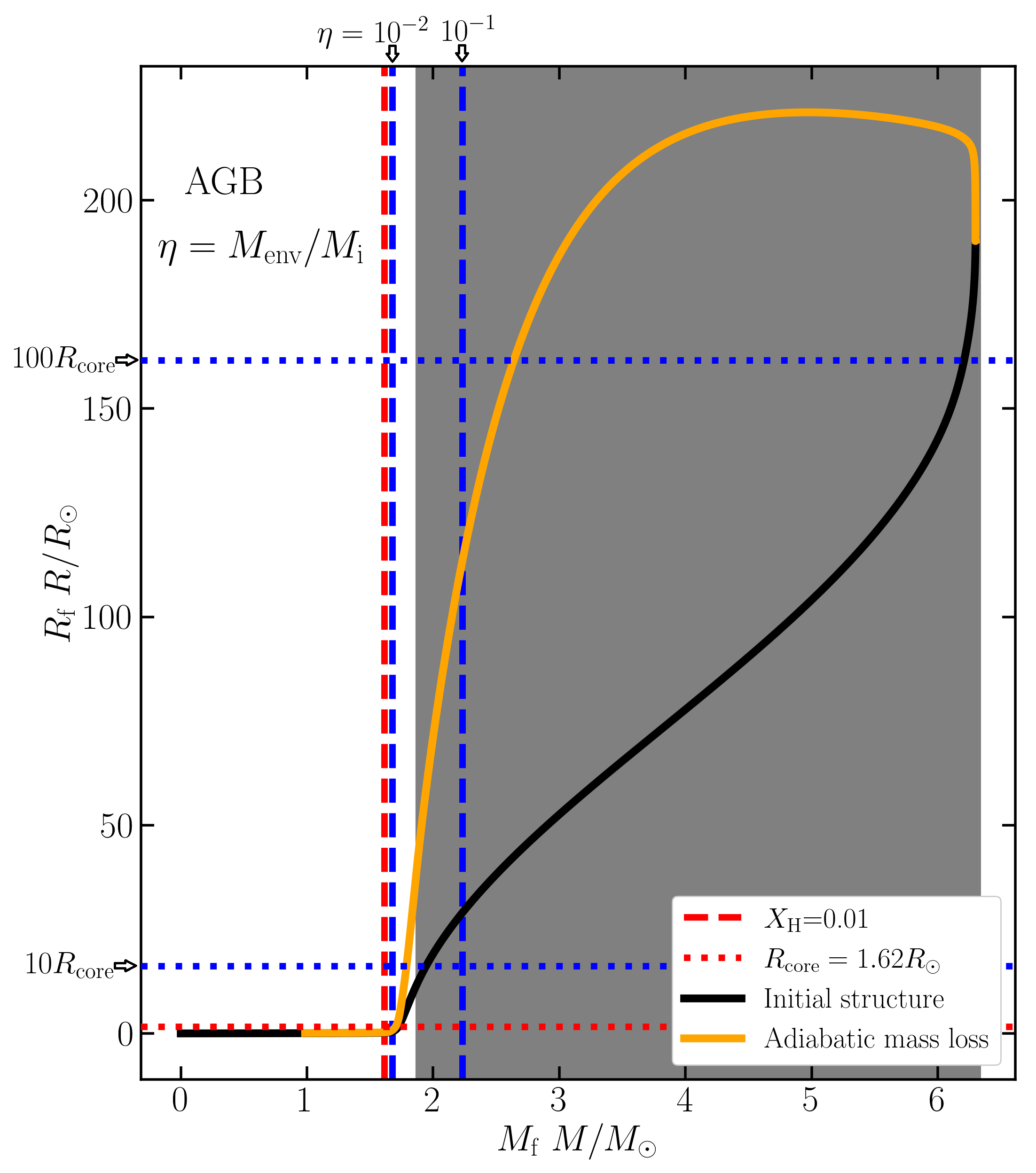}
	\caption{The relationship between the final radius $R_\textrm{f}$ and final mass $M_\textrm{f}$ of an AGB star after CEE. The black solid line is derived from the pre-CEE stellar radius profile, while the orange solid line is obtained from the adiabatic mass loss model. The red dashed line marks the helium core boundary defined by $X_\textrm{H}=0.01$. The blue dashed lines indicate the envelope residual parameters $\eta=0.01$ and $0.1$, and the blue dotted lines correspond to stellar radii of $10\,R_{\odot}$ and $100\,R_{\odot}$.
		\label{Mf-Rf}}
\end{figure}

In Section\,\ref{sec:correction}, we have analysed $\lambda$ as a function of $M_\textrm{env}$ using both the initial stellar structures and adiabatic mass loss models. The results indicate that the changes induced by adiabatic expansion (the $F_2$ effect) are much smaller than those caused by the envelope residual itself. For the vast majority of evolutionary stages, when $M_\textrm{env}<10^{-2}\,M_\textrm{i}$, $F_2$ remains below 1.5. Thus, compared with $F_1$, $F_2$ is relatively unimportant in most stages. However, previous adiabatic mass loss studies (\citealt{PaperIII,2022ApJ...933..137G,2024ApJ...961..202G}) have shown that adiabatic expansion during CEE directly affects the final radius. In Figure\,\ref{Mf-Rf}, we demonstrate the relationship between the final radius $R_\textrm{f}$ and final mass $M_\textrm{f}$ for an AGB star, which shows that adiabatic expansion has a substantial effect on $R_\textrm{f}$. In earlier binary interaction studies, it was not required that the donor remnant fill its Roche lobe at the end of CEE; as a result, the ratio of the donor radius to the Roche-lobe radius is often very small, reaching 0.1 or even 0.01. The envelope residual considered here may offer a solution to this issue. However, \cite{PaperI,2022ApJ...933..137G,2024ApJ...961..202G} suggest that the Roche-lobe filling ratio of the donor star should remain sufficiently high at the end of CEE. In future work, we plan to incorporate the CE energy formalism and the adiabatic mass loss model consistently into BPS codes to obtain self-consistent values of $M_\textrm{f}$ and $R_\textrm{f}$. We have already added the adiabatic $R_\textrm{f}$ to the RBF interpolation grid and uploaded it to our \href{https://github.com/CygX-3-stargate0/Adiabatic_CEE_solver.git}{\textbf{website}}.

In this work, we adopt a solar-metallicity stellar grid without stellar winds. However, the final results in the stellar evolution code can be affected by multiple factors. In the future, low-metallicity grids \citep{2023ApJ...945....7G,PaperV} can be constructed to provide a clearer picture for population II and III stars, and we plan to further investigate the $E_\textrm{bind}$ improvement for other populations.

\section{Summary} \label{summary}

Multiple observations have already indicated that a thin envelope residual should exist after CEE. However, the reduction in $E_\textrm{bind}$ caused by this residual is often neglected. In this work, we have quantified the impact of the envelope residual mass $M_\textrm{env}$ on the CEE binding-energy parameter $\lambda$. In Section\,\ref{sec:method}, we have discussed two independent approaches to determine $\lambda$: one based on the pre-CEE stellar structure (the $F_1$ effect) and the other using the adiabatic mass loss model (the $F_2$ effect). The latter approach can effectively predict the adiabatic expansion during CEE. Our results show that $\lambda$ increases significantly on the RGB and AGB. For example, for a $1\,M_{\odot}$ donor at the RGB tip, an envelope residual of $0.01\,M_{\odot}$ can enhance $\lambda$ by a factor of 15 ($F_1 \sim 15$), with an additional factor of 1.5 from adiabatic expansion ($F_2 \sim 1.5$).

We have tested four different initial settings for the helium-core boundary and $M_\textrm{env}$ (see Table\,\ref{tab:setting}). The results indicate that the dependence of $\lambda$ on the helium-core boundary is weak compared with its dependence on $M_\textrm{env}$. Based on these calculations, we have further constructed RBF interpolation grids and fitting formulae, which we believe will be useful for BPS studies.

We expect that this work will significantly improve the predicted orbital-period distribution of PCEB populations, while reducing the likelihood of CE mergers. It may also help to lower the required $\alpha_\textrm{CE}$ values for forming certain systems, such as short-period sdBs and double WDs.

\begin{acknowledgments}
This project is supported by National Natural Science Foundation of China (NSFC, grant Nos. 12288102, 12525304, 12125303, 12333008, 12422305, 12473033), the Strategic Priority Research Program of the Chinese Academy of Sciences (grant No. XDB1160201), National Key R\&D Program of China (grant Nos. 2021YFA1600403, 2021YFA1600401), Yunnan Revitalization Talent Support Program - Science \& Technology Champion Project (No. 202305AB350003), Yunnan Fundamental Research Projects (No. 202401BC070007), International Centre of Supernovae, Yunnan Key Laboratory (No. 202302AN360001), the CAS "Light of West China" and the Young Talent Project of Yunnan Revitalization Talent Support Program, the New Cornerstone Science Foundation through the XPLORER PRIZE. DH and RC were supported by grant 2024-00123 - Algorithm development for LISA from the Swedish National Space Agency. The data underlying this work are available at \url{https://github.com/CygX-3-stargate0/Adiabatic_CEE_solver.git}. We welcome any comments on this work.
\end{acknowledgments}

\appendix

\section{RBF interpolation on AGB}\label{apd.agb}

\begin{figure}[ht!]
	\plotone{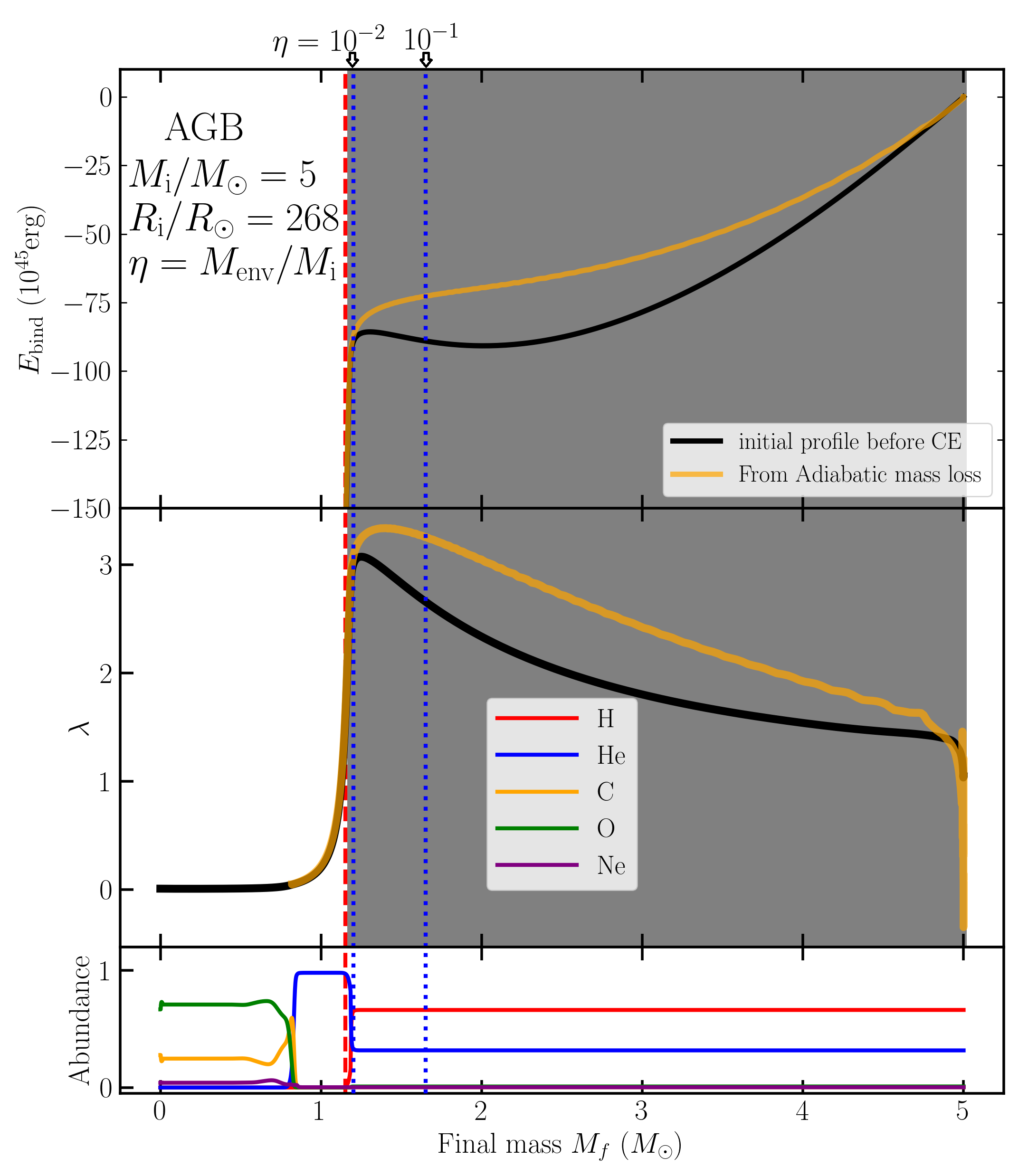}
	\caption{The $E_\textrm{bind}$ and the $\lambda$ as functions of $M_\textrm{f}$ for an AGB star. The black solid lines represent the stellar pre-CEE structure, while the orange solid lines represent the adiabatic mass loss model. The red dashed line marks the helium core boundary defined by $X_\textrm{H}=0.01$. The blue dashed lines indicate the envelope residual parameters $\eta=0.01$ and $0.1$.
		\label{AGB_lam}}
\end{figure}

In Section\,\ref{sec:fit}, we have noted the poor performance of both the RBF interpolation and the polynomial fitting on the AGB branch. This is primarily due to a peak in $\lambda$ located near the core. Figure\,\ref{AGB_lam} provides an example of this behaviour. When $M_\textrm{env}/M_\textrm{i}>0.01$, the gradient of $\lambda$ becomes negative, resulting in a non-monotonic relationship between $\lambda$ and $M_\textrm{env}$. Our results indicate that the RBF interpolation does not handle this type of data well. We therefore recommend avoiding the use of data with $\eta>0.01$ in this regime.

The AGB models exhibiting such structure are found very close to the \textit{super-thermal giant} regime. In fact, as the star evolves further, $M_\textrm{env}$ at the location of the $\lambda$ peak tends to increase to a positive value. Given the complexity of the \textit{super-thermal giant} phase, a more detailed investigation using alternative assumptions is required, and we do not pursue this issue further in the present work.

\bibliography{sample702}{}
\bibliographystyle{aasjournalv7.1}



\end{document}